\documentclass[prd,twocolumn,showpacs,preprintnumbers,amsmath,amssymb,superscriptaddress]{revtex4}
\usepackage{graphicx}
\usepackage{amsmath}
\usepackage{amssymb}
\usepackage{afterpage}
\usepackage{graphics}
\usepackage{float}
\usepackage{rotating}
\usepackage{xspace}
\usepackage{epsfig}
\usepackage{dcolumn}
\usepackage{bm}
\usepackage{multirow}
\usepackage{hhline}

\newcommand{\nue}{\nu_{e}}
\newcommand{\num}{\nu_{\mu}}
\newcommand{\nut}{\nu_{\tau}}

\newcommand{\numbar}{\overline{\nu}_{\mu}}
\newcommand{\nutbar}{\overline{\nu}_{\tau}}

\newcommand{\sinSqr}{\sin^{2}2\theta_{23}}
\newcommand{\deltaMSqr}{\Delta m^{2}_{23}}

\newcommand{\mup}{\mu^{+}}
\newcommand{\mum}{\mu^{-}}

\newcommand{\chiLine}{\chi^{2}_{\text{line}}/ndf}
\newcommand{\lowRes}{\text{unknown momentum}}
\newcommand{\lowMom}{\text{low momentum}}
\newcommand{\hiMom}{\text{high momentum}}
\newcommand{\liveTime}{854.24}
\newcommand{\acceptance}{6.91\times10^{6}\text{cm}^{2}\text{sr}}
\newcommand{\candidates}{140}
\newcommand{\upgoing}{130}
\newcommand{\horizontal}{10}
\newcommand{\nullExclude}{94}
\newcommand{\lowHi}{0.65}
\newcommand{\lowHiUp}{0.15}
\newcommand{\lowHiDn}{0.12}
\newcommand{\lowHiM}{0.49}
\newcommand{\lowHiMUp}{0.20}
\newcommand{\lowHiMDn}{0.14}
\newcommand{\lowHiP}{0.55}
\newcommand{\lowHiPUp}{0.30}
\newcommand{\lowHiPDn}{0.19}
\newcommand{\mtop}{0.72}
\newcommand{\mtopUp}{0.24}
\newcommand{\mtopDn}{0.18}
\newcommand{\mtopSys}{^{+0.08}_{-0.04}}
\newcommand{\cpt}{0.89}
\newcommand{\cptUp}{0.54}
\newcommand{\cptDn}{0.33}
\newcommand{\cptSys}{0.03}

\begin{document}
\preprint{FERMILAB-PUB-07-012-E, BNL-77481-2007-JA, hep-ex 0701045}

\title{Charge-Separated Atmospheric Neutrino-Induced Muons in the MINOS Far Detector}                              

\newcommand{\Cambridge}{Cavendish Laboratory, University of Cambridge, Madingley Road, Cambridge CB3 0HE, United Kingdom}
\newcommand{\FNAL}{Fermi National Accelerator Laboratory, Batavia, Illinois 60510, USA}
\newcommand{\RAL}{Rutherford Appleton Laboratory, Chilton, Didcot, Oxfordshire, OX11 0QX, United Kingdom}
\newcommand{\UCL}{Department of Physics and Astronomy, University College London, Gower Street, London WC1E 6BT, United Kingdom}
\newcommand{\Caltech}{Lauritsen Laboratory, California Institute of Technology, Pasadena, California 91125, USA}
\newcommand{\ANL}{Argonne National Laboratory, Argonne, Illinois 60439, USA}
\newcommand{\Athens}{Department of Physics, University of Athens, GR-15771 Athens, Greece}
\newcommand{\NTUAthens}{Department of Physics, National Tech. University of Athens, GR-15780 Athens, Greece}
\newcommand{\Benedictine}{Physics Department, Benedictine University, Lisle, Illinois 60532, USA}
\newcommand{\BNL}{Brookhaven National Laboratory, Upton, New York 11973, USA}
\newcommand{\CdF}{APC -- Coll\`{e}ge de France, 11 Place Marcelin Berthelot, F-75231 Paris Cedex 05, France}
\newcommand{\Cleveland}{Cleveland Clinic, Cleveland, Ohio 44195, USA}
\newcommand{\GEHealth}{GE Healthcare, Florence South Carolina 29501, USA}
\newcommand{\Harvard}{Department of Physics, Harvard University, Cambridge, Massachusetts 02138, USA}
\newcommand{\HolyCross}{Holy Cross College, Notre Dame, Indiana 46556, USA}
\newcommand{\IIT}{Physics Division, Illinois Institute of Technology, Chicago, Illinois 60616, USA}
\newcommand{\Indiana}{Physics Department, Indiana University, Bloomington, Indiana 47405, USA}
\newcommand{\ITEP}{High Energy Experimental Physics Department, Institute of Theoretical and Experimental Physics, 
  B. Cheremushkinskaya, 25, 117218 Moscow, Russia}
\newcommand{\JMU}{Physics Department, James Madison University, Harrisonburg, Virginia 22807, USA}
\newcommand{\LASL}{Nuclear Nonproliferation Division, Threat Reduction Directorate, Los Alamos National Laboratory, Los Alamos, New Mexico 87545, USA}
\newcommand{\Lebedev}{Nuclear Physics Department, Lebedev Physical Institute, Leninsky Prospect 53, 117924 Moscow, Russia}
\newcommand{\LLL}{Lawrence Livermore National Laboratory, Livermore, California 94550, USA}
\newcommand{\MIT}{Lincoln Laboratory, Massachusetts Institute of Technology, Lexington, Massachusetts 02420, USA}
\newcommand{\Minnesota}{University of Minnesota, Minneapolis, Minnesota 55455, USA}
\newcommand{\Crookston}{Math, Science and Technology Department, University of Minnesota -- Crookston, Crookston, Minnesota 56716, USA}
\newcommand{\Duluth}{Department of Physics, University of Minnesota -- Duluth, Duluth, Minnesota 55812, USA}
\newcommand{\Oxford}{Subdepartment of Particle Physics, University of Oxford,  Denys Wilkinson Building, Keble Road, Oxford OX1 3RH, United Kingdom}
\newcommand{\Pittsburgh}{Department of Physics and Astronomy, University of Pittsburgh, Pittsburgh, Pennsylvania 15260, USA}
\newcommand{\IHEP}{Institute for High Energy Physics, Protvino, Moscow Region RU-140284, Russia}
\newcommand{\RoyalH}{Physics Department, Royal Holloway, University of London, Egham, Surrey, TW20 0EX, United Kingdom}
\newcommand{\Carolina}{Department of Physics and Astronomy, University of South Carolina, Columbia, South Carolina 29208, USA}
\newcommand{\SLAC}{Stanford Linear Accelerator Center, Stanford, California 94309, USA}
\newcommand{\Stanford}{Department of Physics, Stanford University, Stanford, California 94305, USA}
\newcommand{\Sussex}{Department of Physics and Astronomy, University of Sussex, Falmer, Brighton BN1 9QH, United Kingdom}
\newcommand{\TexasAM}{Physics Department, Texas A\&M University, College Station, Texas 77843, USA}
\newcommand{\Texas}{Department of Physics, University of Texas, 1 University Station, Austin, Texas 78712, USA}
\newcommand{\TechX}{Tech-X Corporation, Boulder, Colorado 80303, USA}
\newcommand{\Tufts}{Physics Department, Tufts University, Medford, Massachusetts 02155, USA}
\newcommand{\UNICAMP}{Universidade Estadual de Campinas, IF-UNICAMP, CP 6165, 13083-970, Campinas, SP, Brazil}
\newcommand{\USP}{Instituto de F\'{i}sica, Universidade de S\~{a}o Paulo,  CP 66318, 05315-970, S\~{a}o Paulo, SP, Brazil}
\newcommand{\Washington}{Physics Department, Western Washington University, Bellingham, Washington 98225, USA}
\newcommand{\WandM}{Department of Physics, College of William \& Mary, Williamsburg, Virginia 23187, USA}
\newcommand{\Wisconsin}{Physics Department, University of Wisconsin, Madison, Wisconsin 53706, USA}
\newcommand{\deceased}{Deceased.}

\affiliation{\ANL}
\affiliation{\Athens}
\affiliation{\Benedictine}
\affiliation{\BNL}
\affiliation{\Caltech}
\affiliation{\Cambridge}
\affiliation{\UNICAMP}
\affiliation{\CdF}
\affiliation{\FNAL}
\affiliation{\Harvard}
\affiliation{\IIT}
\affiliation{\Indiana}
\affiliation{\IHEP}
\affiliation{\ITEP}
\affiliation{\JMU}
\affiliation{\Lebedev}
\affiliation{\LLL}
\affiliation{\UCL}
\affiliation{\Minnesota}
\affiliation{\Duluth}
\affiliation{\Oxford}
\affiliation{\Pittsburgh}
\affiliation{\RAL}
\affiliation{\USP}
\affiliation{\Carolina}
\affiliation{\Stanford}
\affiliation{\Sussex}
\affiliation{\TexasAM}
\affiliation{\Texas}
\affiliation{\Tufts}
\affiliation{\Washington}
\affiliation{\WandM}
\affiliation{\Wisconsin}

\author{P.~Adamson}
\affiliation{\FNAL}
\affiliation{\UCL}

\author{C.~Andreopoulos}
\affiliation{\RAL}

\author{K.~E.~Arms}
\affiliation{\Minnesota}

\author{R.~Armstrong}
\affiliation{\Indiana}

\author{D.~J.~Auty}
\affiliation{\Sussex}

\author{S.~Avvakumov}
\affiliation{\Stanford}

\author{D.~S.~Ayres}
\affiliation{\ANL}

\author{B.~Baller}
\affiliation{\FNAL}

\author{B.~Barish}
\affiliation{\Caltech}

\author{P.~D.~Barnes~Jr.}
\affiliation{\LLL}

% ...

\collaboration{The MINOS Collaboration}
\noaffiliation

\date{\today}                                           

\begin{abstract}
We found $\candidates$ neutrino-induced muons in $\liveTime$ live days in the MINOS far detector, which has an acceptance for neutrino-induced muons of $\acceptance$.  We looked for evidence of neutrino disappearance in this data set by computing the ratio of the number of $\lowMom$ muons to the sum of the number of $\hiMom$ and $\lowRes$ muons for both data and Monte Carlo expectation in the absence of neutrino oscillations.  The ratio of data and Monte Carlo ratios, $\cal{R}$, is 
\begin{equation}
{\cal R} =\lowHi^{+\lowHiUp}_{-\lowHiDn}(\text{stat})\pm0.09(\text{syst}), 
\nonumber
\end{equation}
a result that is consistent with an oscillation signal. A fit to the data for the oscillation parameters $\sinSqr$ and $\deltaMSqr$ excludes the null oscillation hypothesis at the $\nullExclude$\% confidence level.  We separated the muons into $\mum$ and $\mup$ in both the data and Monte Carlo events and found the ratio of the total number of $\mum$ to $\mup$ in both samples.  The ratio of those ratios, $\hat{\mathcal{R}}_{CPT}$, is a test of CPT conservation.  The result 
\begin{equation}
\hat{\mathcal{R}}_{CPT} =  \mtop^{+\mtopUp}_{-\mtopDn}(\text{stat})\mtopSys(\text{syst}),
\nonumber
\end{equation}
is consistent with CPT conservation.  
\end{abstract}

\maketitle

\section{\label{sec:intro}Introduction}     

Measurements of atmospheric neutrinos by Super-Kamiokande experiment have shown that there is a deficit of $\num$ when compared to expectations~\cite{Fukuda1, Fukuda2, Fukuda3, Fukuda4, Fukuda6, Fukuda5, Ashie:2004sm, Ashie:2005ik, Abe:2006fu}.
The hypothesis that best describes this deficit is the oscillation of $\num(\numbar)\rightarrow\nu_{\tau}(\nutbar)$~ \cite{PDBook,Kayser:2005cd}, with the oscillation probability given by
\begin{equation}
P_{\num\rightarrow\nu_{\tau}} = \sinSqr\sin^{2}(1.27\deltaMSqr L/E),
\label{eq:standardprob}
\end{equation} 
where $\theta_{23}$ is the mixing angle, $\Delta m^{2}_{23} = |m_3^2-m_2^2| $ is the mass squared difference in eV$^{2}$ between the neutrino mass states, $L$ is the distance in km traveled by the neutrino, or its baseline, and $E$ is the energy of the $\num$ in GeV~\cite{final}. The Super-Kamiokande data is best fit by the oscillation hypothesis with parameters $(\sinSqr, \deltaMSqr)=(1.0,2.4 \times 10^{-3}\,{\rm eV^2})$ \cite{Ashie:2005ik, Abe:2006fu}; the ranges for these parameters given by the 90\% confidence contours of the zenith angle oscillation fit are $\sinSqr > 0.92$ and $1.5 <\deltaMSqr < 3.4 \times 10^{-3}\,{\rm eV^2}$.  The MACRO~\cite{Ambrosio:2001je, Ambrosio:2004sm} and Soudan 2~\cite{Sanchez:2003rb, Allison:2005dt} results are consistent with those obtained by Super-Kamiokande.  For the MINOS analysis of atmospheric neutrinos with an interaction vertex in the detector,  the parameter ranges are $\sinSqr > 0.2$ and $7 \times 10^{-5} < \deltaMSqr < 5 \times 10^{-2}\,{\rm eV^2}$~\cite{Adamson:2005qc}.  Below we extend the MINOS atmospheric analysis to neutrino-induced muons observed in the MINOS far detector.

The oscillation hypothesis for the atmospheric neutrino deficit has received strong support from the first results of the MINOS long baseline experiment.   MINOS, which sends $\num$ produced at Fermilab to a detector 735~km away in northern Minnesota \cite{minosD}, finds  $(\sinSqr, \deltaMSqr)=(1.00,2.74 \times 10^{-3}\,{\rm eV^2})$ \cite{MINOSnu}, when fitting their $\nu_\mu$ events; the ranges of these parameters given by the 90\% confidence contours are $\sinSqr > 0.72$ and $2.2 < \deltaMSqr < 3.8 \times 10^{-3}\,{\rm eV^2}$.   The K2K long baseline experiment, which sent $\num$ produced at KEK to the Super-Kamiokande detector located 250~km away, also found consistent results --  $(1.0,2.8 \times 10^{-3}\,{\rm eV^2})$ \cite{Aliu:2004sq, ahn} with 90\% confidence ranges for the parameters of  $\sinSqr > 0.55$ and $1.9 < \deltaMSqr < 3.6 \times 10^{-3}\,{\rm eV^2}$.

Since the magnetized MINOS far detector distinguishes $\mu^-$ from $\mu^+$, MINOS data can be used as a probe of CPT conservation in the neutrino sector.  CPT conservation requires that the probability of an atmospheric $\num$ of a given energy remaining a $\num$ after traveling from its origin to its point of interaction be equal to the probability that an atmospheric $\numbar$ of the same energy remains a $\numbar$ after traveling the same distance~\cite{Kayser:2005cd}.  We consequently test CPT conservation by comparing $\num$-induced $\mu^-$ and $\numbar$-induced $\mu^+$ with respect to expectations, as a measure of whether the atmospheric $\num$ and $\numbar$ are disappearing at the same rate.  Effects induced by charged current interactions of the $\nue$ components of the neutrino eigenstates with the earth could masquerade as apparent CPT violation.  However, distortions from matter effects are estimated to be small for most of the $\theta_{13}$ range compatible with the CHOOZ limit~\cite{Apollonio:1999ae}. The average induced $\num/\numbar$ event rate asymmetry is estimated to be less than 10\%~\cite{Palomares-Ruiz:2004tk}. Therefore, at our current level of statistical accuracy we can ignore matter effects and describe $\num(\numbar)\rightarrow\nut(\nutbar)$ oscillations using the two flavor approximation in Eq.~(\ref{eq:standardprob}).

After briefly discussing the detector in \S\ref{sec:det}, we describe our event sample in \S\ref{sec:dataSample} and the Monte Carlo event generation in \S\ref{sec:sim}.  In \S\ref{sec:upSelection}, we describe the cuts used in selecting the sample of  muons to be analyzed.  We present the oscillation analysis in \S\ref{sec:results} and the probe of CPT conservation using charge separated neutrino-induced $\mum$ and $\mup$ in \S\ref{sec:CPT}.

\section{ \label{sec:det}The MINOS Far Detector}

The MINOS far detector is a steel-scintillator sampling calorimeter located at a depth of 2070 meters-water-equivalent (m.w.e.) in the Soudan mine in northern Minnesota~\cite{minosD}.  The detector is made up of 486 vertical octagonal planes of 2.54~cm thick steel laminates, interleaved with 484 planes of 1~cm thick extruded polystyrene scintillator strips and a 2.5~cm wide air gap.  Each scintillator plane has 192 strips of width 4.1~cm.  The length of each strip depends on its position in the plane and varies between 3.4~and~8.0~m.  The scintillator strips in alternating detector planes are oriented at $\pm 45^\circ$ to the vertical. The modular detector consists of two supermodules (SM) separated by a gap of 1.1~m.  The detector coordinate system has the {\it y} axis pointing up, the {\it z} axis along the detector long axis, pointing away from Fermilab and the {\it x} axis forms a right-handed coordinate system.

Scintillation light from charged particles traversing the MINOS plastic scintillator is collected with wavelength shifting (WLS) plastic fibers embedded in the scintillator strips.  The WLS fibers are coupled to clear optical fibers at both ends of a strip and are read out using 16-pixel multi-anode phototmultiplier tubes (PMTs).  The signals from eight strips, separated by approximately 1 m within the same plane, are optically summed, or multiplexed, and read out by a single PMT pixel.  The multiplexing pattern is different for the two sides of the detector, enabling the resulting eightfold ambiguity to be resolved for single particles.  For all other types of events, ambiguities are resolved effectively using additional information from timing and event topology.  

To measure the momentum of muons traversing the detector, the steel has been magnetized into a toroidal field configuration using a current-carrying coil that runs through the central axis of each SM.  A finite element analysis calculation shows that each SM is magnetized to an average value of $\sim1.3$~T by the 15~kA current loop.  The field is saturated near the coil hole at a strength of approximately 1.8~T, falling to about 1~T near the edges.  Further details about the MINOS detector can be found in \cite{minosD}.

\begin{figure}
\centerline{\epsfig{file=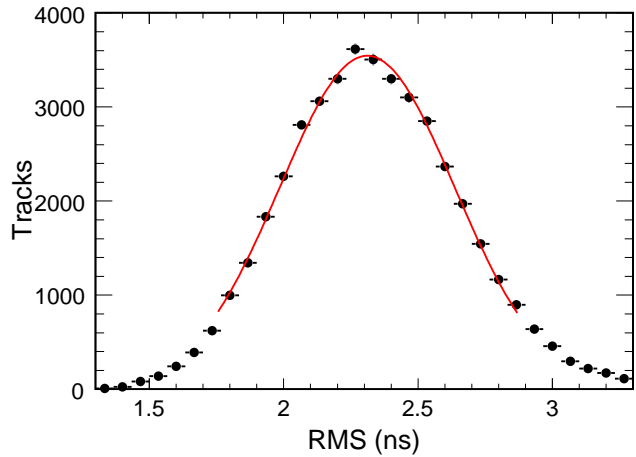, width=3.75in}}
\caption[Distribution of timing constant rms]{Distribution of the r.m.s.~deviations of the measured hit times from the calculated hit times for calibration tracks.  The resolution of the system is the mean of a Gaussian fit to the distribution, $2.31\pm0.03$~ns.}
\label{fig:timing_res}
\end{figure}

To distinguish upward-going neutrino-induced muons from the downward cosmic ray background requires excellent timing resolution.  The MINOS timing system has a least count of 1.56~ns.  However, the timing resolution is dominated by the 8~ns fluorescence decay time in the WLS fiber~\cite{Border:2001vk}. The overall resolution is determined by the convolution of the least count
with the fluctuations in the arrival times of the photons at the PMT.  We calibrated the timing system by measuring the time offsets between each channel.  For this determination~\cite{Rebel,Blake}, we selected a sample of downward through-going cosmic ray muons with well-reconstructed tracks.  For each hit along a track we measured the travel times from the track entrance point in the detector to the hit location and corrected those values for both the rise time of the signal and the propagation of light along the fibers.  We then compared these times with the time expected for a relativistic muon traversing the same distance and computed the offset between the measured and expected times for each hit, $\Delta t$.  The timing calibration constants are obtained using an iterative method to minimize the mean $\Delta t$ for each channel.   After calibration, linear timing fits are again applied to the times and positions of the hits on each muon track. The r.m.s.~deviation between the measured and fitted times is calculated for each track. Figure~\ref{fig:timing_res} shows the distribution of these r.m.s.~deviations. In this figure, the distribution has been fit to a Gaussian with a mean of $2.31\pm0.03$~ns, a value that represents the overall timing resolution of the detector.

\section{ \label{sec:data} Data Analysis}

In the analyses presented here we use neutrino-induced muons, which are defined as events that come from below or slightly above the horizon.  These events are essentially uncontaminated by the background of downward-going atmospheric muons.  We analyze two types of events:  those that pass completely through  the detector (``through-going muons'') and those that enter and stop in the detector (``stopping muons'').  Earlier MINOS results for events with an interaction vertex in the detector (``contained events'') appear in \cite{Adamson:2005qc}. We distinguish between neutrino-induced muons with well-determined momenta and neutrino-induced muons whose charge and momentum are undetermined~(c.f.,~\S\ref{sec:charge}). 

\subsection{ \label{sec:dataSample}Data Sample}

The data presented here represent $\liveTime$ live days and were taken between August 1, 2003 and April 30, 2006.   The geometric acceptance of the detector for neutrino-induced muons is $\acceptance$.  We found a total of $\candidates$ neutrino-induced muons in this data set.  

\subsection{\label{sec:sim} Simulated Muons}
\subsubsection{Atmospheric Neutrino-induced Muons}

\begin{figure}
\centerline{\epsfig{file=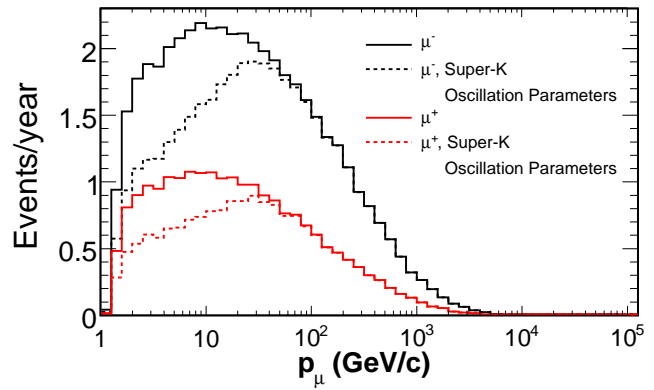, width=3.75in}}
\caption[Distribution of momentum values for $\mum$ and $\mup$.]{Input momentum distribution for neutrino-induced $\mum$ and $\mup$ calculated with the {\scshape Nuance} simulation package.  The distributions in the MINOS Far detector without oscillations and with the Super-Kamiokande determined values of $\sinSqr$ and $\deltaMSqr$ are shown. The oscillation affects mostly muons with low momenta.}
\label{fig:nuancemom}
\end{figure}

We generated a large sample of simulated neutrino-induced muons in the MINOS far detector using the Bartol~96~\cite{Agrawal:1996gk} neutrino flux tables along with the default {\scshape Nuance} neutrino interaction model~\cite{Casper:2002sd} and the {\scshape Grv94}~\cite{Gluck:1994uf} parton distribution functions. We first simulated neutrino-induced muons passing through the surface of a box surrounding the MINOS far detector and then propagated these muons through the detector with the MINOS {\scshape Geant}-based detector Monte Carlo simulation.  A total of $6.5\times10^{5}$ neutrino-induced muon events, the equivalent of 2500 years of live time, were generated in this way.  The momentum distributions simulated for the neutrino-induced $\mum$ and $\mup$ in MINOS are shown in Fig.~\ref{fig:nuancemom} for neutrinos without oscillations and for neutrinos with an oscillation signal using the Super-Kamiokande parameters~\cite{Ashie:2005ik}.  This figure shows that the oscillation signal affects mostly muons with low momenta.

\subsubsection{Cosmic Ray Muons}

We simulated each cosmic ray muon event by first choosing its arrival direction from a distribution uniform in solid angle and then associating this direction with the overburden found in the Soudan 2 slant depth map~\cite{Kasahara:1997kw}.  The surface energy of the muon was selected from the known distribution~\cite{part}.  Once the surface energy and overburden were chosen, we tested the muon to see whether it survived energy loss processes between the surface and the detector~\cite{part}.  We placed the muons that survived on an imaginary box surrounding the detector~\cite{Hawthorne:1998mz} and then propagated them though the detector with the MINOS GEANT-based detector simulation.  

\subsection{ \label{sec:upSelection}Event Selection}

All data and simulated muons were analyzed using the standard MINOS reconstruction algorithms and a uniform set of event selection cuts.  The selection cuts summarized below are  described in further detail in~\cite{Rebel}.

\subsubsection{\label{sec:mu}Muon Selection Cuts}

We first selected muon events using criteria developed for the study of cosmic ray muons with the MINOS far detector~\cite{m&r}. The first two cuts require that there was a track successfully reconstructed in the event (``No Track'' cut), and that there is only a single track found by the track-fitting algorithm (``Multiples'' cut).  The next set of cuts exclude random collections of hits that could be mistaken for muon tracks.  These cuts require that: the track must cross at least 20 planes in the detector (``20 Plane'' cut); the track must have a pathlength of at least 2~m (``2.0~m Length'' cut); the earliest recorded hit of the track be no more than 15~cm from the front or back surface of the detector or 30~cm from any of the remaining surfaces (``Fiducial'' cut); and the track fit must have a $\chi^{2}$ per number of degrees of freedom, $\chi^{2}_{fit}/ndf<1.5$.  

\begin{figure}
\centerline{\epsfig{file=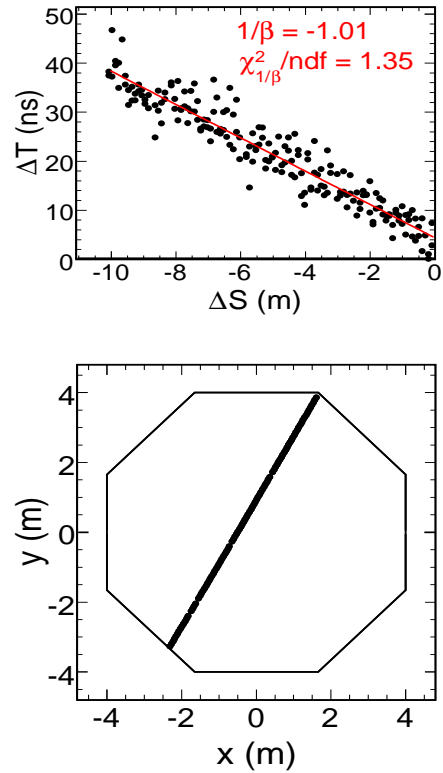, width=2.5in, height=4.25in}}
\caption{A typical cosmic ray muon in the MINOS far detector.  The top panel shows the timing information for the hits along the track with a straight line fit superposed. The legend gives $\Delta T/\Delta S = 1/\beta$ and the $\chi^{2}_{1/\beta}/ndf$ of the fit for this muon. The bottom panel shows the $(x,y)$ hit positions of this track.  The resolution of the positions is the 4~cm width of the strips.}
\label{fig:downEvent}
\end{figure}

The next selection cuts were designed to remove events with poor timing information.  To ensure the presence of sufficient timing information in the events, we first excluded tracks if fewer than half of the hits come from strips with signals on both ends (``Double-ended Strip Cut").  We determined the direction of the track by plotting the time difference $\Delta T$~(ns) of each hit along the track as a function of its distance $\Delta S$~(m) from the first hit.  If the $\it y$ positions of the hits increase along the length of the track, $\Delta S$ is positive; for $\it y$ decreasing along the track, $\Delta S$ is negative.  The slope of the linear fit to the $\Delta T/\Delta S$ distribution is $1/\beta = c/v$.  The top panel of Fig.~\ref{fig:downEvent} shows the $\Delta T/\Delta S$ distribution with the linear fit superposed for a typical cosmic muon; the bottom panel displays the $(x,y)$ positions of the hits in the detector for this muon.  

\begin{figure}
\centerline{\epsfig{file=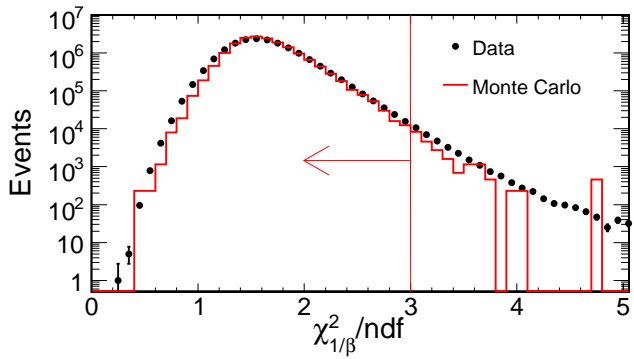, width=3.75in}}
\caption[Distribution of $\chi^{2}/ndf$ values from $1/\beta$ determination.]{Distribution of 
$\chi^{2}_{1/\beta}/ndf$ values from the $1/\beta$ determinations for data and cosmic ray muon Monte Carlo events.  In this figure, the Monte Carlo distribution has been normalized to the same area as the data distribution. Events with $\chi^{2}_{1/\beta}/ndf < 3$ are used in the analysis.}
\label{fig:invbetachi2}
\end{figure}

In Fig.~\ref{fig:invbetachi2} we show the distributions of $\chi_{1/\beta}^{2}/ndf$ values for the linear fits to $1/\beta$ for both data and cosmic ray muon Monte Carlo events passing the above cuts.  The distribution does not peak at 1 as might be expected for a $\chi^{2}/ndf$ statistic.  However, that is unimportant to the analysis since we are using the $\chi_{1/\beta}^{2}/ndf$ value only to define a cut.  We defined a cut that requires $\chi_{1/\beta}^{2}/ndf<3.0$ for an event (``$\chi_{1/\beta}^{2}/ndf<3.0$'' cut).  The cut was selected at the value of $\chi_{1/\beta}^{2}/ndf$ where the number of events falls to $\sim$1\% of the peak.  This cut maximizes the number of selected events while excluding those events with the worst $\chi_{1/\beta}^{2}/ndf$ values.  Although the data and Monte Carlo simulation deviate at low values of $\chi_{1/\beta}^{2}/ndf$ the agreement is excellent near the peak and the cut value. Therefore, the Monte Carlo sample can be used to effectively study the systematic uncertainty introduced into the analysis by this cut.

In the last step of the muon selection, we separated the downward-going from upward-going events.  Upward-going events have a positive slope for the straight line fit to the $\Delta T/\Delta S$ distribution.  The final muon cut is a check on the up/down separation.  The entrance point of a track, and therefore its incoming direction, is determined by the slope of the times of the hits in the detector as a function of their {\it z} position.  For a few short tracks this information is sufficiently ambiguous that the entrance point can be confused with the exit point of the track causing the reconstruction to interchange the two.  As a result, downward-going muons are incorrectly reconstructed as upward-going.  We remove these failures by checking the times of the hits in the reconstructed track as a function of their $\it y$ position.  The slope of the line fit to the hits as a function of their $\it y$  positions must agree with the reconstructed incoming direction of the track (``Directionality'' cut). 

\begin{table}
\caption{\label{table:selectioncuts}Effect of cuts on data and simulated samples.  The fraction of the total events remaining is shown.  $N$ gives the total number of events in each sample. }
\begin{ruledtabular}
\begin{tabular}{|l|c|c|c|} 
\multicolumn{1}{|c|}{Cut} & Data & $\mu$ MC& $\nu$-induced MC\\ \hline
\hline & $N=$ &$N=$&$N=$\\
 & ~$3.81\times10^{7}$~&~$1.20\times10^{6}$~&$6.47\times10^{5}$\\ %
\hline
\hline No Cuts                             & 1.000    & --       & 1.000 \\ 
\hline No Track                            & 0.800   & --       & 0.862 \\ 
\hline Multiples                             & 0.743   & --       & 0.848 \\ 
\hline 20 Plane                              & 0.561   & 0.561  & 0.606\\ 
\hline 2.0~m Length                      & 0.557   & 0.557  & 0.578 \\ 
\hline Fiducial                               & 0.534   & 0.538  & 0.559 \\ 
\hline $\chi^{2}_{fit}/ndf<1.5$       & 0.429    & 0.447  & 0.497 \\ 
\hline Double-ended Strips            & 0.429   & 0.447  & 0.497 \\ 
\hline $\chi^{2}_{1/\beta}/ndf<3.0$& 0.428    & 0.447  & 0.497 \\ 
\hline Directionality                      & 0.428    & 0.447 & 0.478 \\ 
\end{tabular}
\end{ruledtabular}
\end{table}

\begin{figure}
\centerline{\epsfig{file=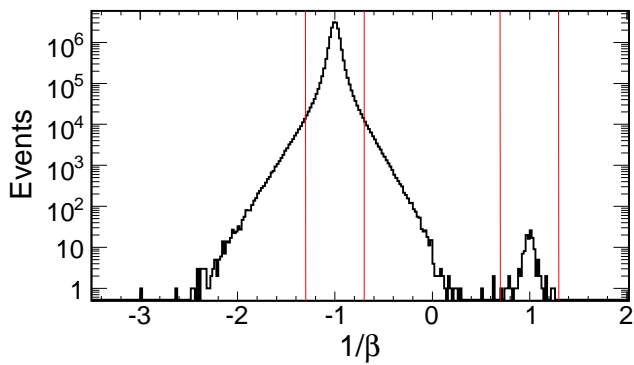, width=3.75in, width=3.75in}}
\caption[Distribution of $1/\beta$ for upward-going and cosmic ray muons.]{Distribution of $1/\beta$ for upward-going neutrino-induced muons, with a peak at 1,  and downward-going cosmic ray muons, with a peak at -1. The vertical lines at 0.7 and 1.3 bracket the events included in the upward-going muon sample.}
\label{fig:invbeta}
\end{figure}

Table~\ref{table:selectioncuts} shows the effect of the cuts on the data and the two Monte Carlo distributions.  We normalize the cosmic ray muon Monte Carlo distributions to the data value at the 20 Plane cut because the Monte Carlo simulation does not account for multiple muon events which make up $\sim 5\%$ of downward-going cosmic ray muons~\cite{gaisser}.  After cuts the selected events consist of muon tracks whose directionality is well determined.  The distribution of $1/\beta$ values for the selected muon events is shown in Fig.~\ref{fig:invbeta}.  Since the sample includes both downward-going and upward-going muons and these muons are relativistic, there are two peaks in the $1/\beta$ distribution.  The peak at $1/\beta = -1$ are downward-going muons and the peak at $1/\beta = 1$ are upward-going muons.  Integrating the events in the two peaks shows that the fraction of (upward-going muons)/(downward-going muons) $\sim 10^{-5}$.

\subsubsection{\label{sec:up}Upward-Going Neutrino-Induced Muons}

\begin{figure}
\centerline{\epsfig{file=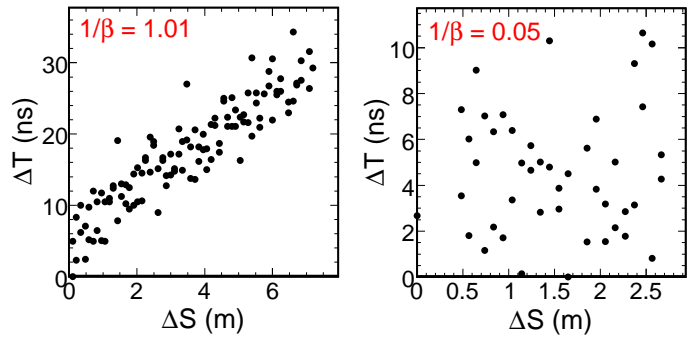, width=3.75in}}
\caption{Distributions of $\Delta T/\Delta S$ as in Fig.~\ref{fig:downEvent}.  The left panels shows a typical upward-going muon and the right panel shows a muon excluded by the $1/\beta$ cut from the upward-going sample.}
\label{fig:upMus}
\end{figure}

As in previous experiments~\cite{Fukuda4,Allison:2005dt}, we identify upward-going muons that enter the detector as neutrino-induced muons.  Events in the range $0.7<1/\beta<1.3$ are included in our neutrino-induced muon data set.  The cut was determined using downward-going cosmic ray muons.  It was set as the $1/\beta$ range that includes 99\% of the events and  which is centered on the peak.  In the left panel of Fig.~\ref{fig:upMus} we show a well-characterized upward-going muon event.  The right panel shows an event excluded by the $1/\beta$ cut. 

Table~\ref{table:upHorizontal} shows that our data sample includes $\upgoing$ upward-going, neutrino-induced muons in the range $0.7<1/\beta<1.3$.  In this table we also show the  background from contained vertex interactions where the neutrino interacts close to the detector edge and cannot be separated from muons entering the detector.  This background was estimated by calculating the rate of Monte Carlo atmospheric neutrino events that interact inside the detector volume~\cite{Adamson:2005qc}.  The background includes events that pass the cuts in Table~\ref{table:selectioncuts} and satisfy the upward-going selection cut.  The Monte Carlo events were assumed to oscillate with the best-fit oscillation parameters from Super-Kamiokande~\cite{Ashie:2005ik}.   The 20 Plane cut has been shown to be effective at removing the background due to upward-going pions produced by downward-going cosmic ray muons interacting in the rock surrounding the detector.

\subsubsection{\label{sec:horizontal}Horizontal Muons}

In this analysis, we also include muons coming from along and slightly above the detector's horizon.  The flat overburden of the Soudan site~\cite{Kasahara:1997kw} makes this search feasible. The slant depth of rock between the detector and the surface for incoming directions above the horizon increases approximately as $\sec\theta$, where $\theta$ is the zenith angle.  Only the highest energy cosmic ray muons ($E_{\mu} > 100$~TeV) have sufficient energy to penetrate the large column of rock present for muons coming from directions near the horizon.  Since the intensity of cosmic ray muons falls as $E^{-2.7}$, very few cosmic ray muons survive to reach the detector, implying that muons from near the detector's horizon are neutrino-induced.  These events are important to the analysis because they sample neutrino-induced muons with lower values of $L/E$ than the upward-going muons.  Eq.~(\ref{eq:standardprob}) shows that the average oscillation probability for these horizontal muons is lower than for upward-going neutrino-induced muons and they add information important in determining the neutrino flux normalization, a parameter used in the oscillation analysis in \S\ref{sec:oscFit}.  

\begin{figure}
\centerline{\epsfig{file=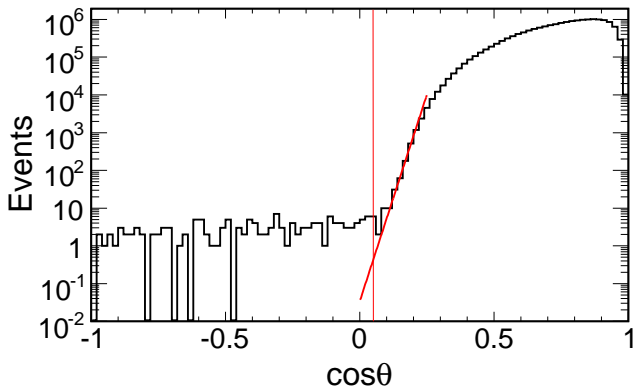, width=3.75in}}
\caption{The $\cos\theta$ distribution for all incoming muons. The data are fit to an exponential between $0.1 \leq \cos\theta \leq 0.2$ to estimate the background of cosmic ray muons in the horizontal neutrino-induced muon signal region. The fit is shown in the figure.  As indicated, we select muons coming from directions with $\cos\theta < 0.05$.  }
\label{fig:costheta}
\end{figure}

We require that the measured velocity of the downward-going muons be in the range $-1.3<1/\beta<-0.7$ .  To determine the maximum angle above the horizon from which we can distinguish horizontal muons from cosmic ray muons, we use the zenith angle distribution for all events shown in Fig.~\ref{fig:costheta}.  The distribution is steeply falling for $\cos\theta < 0.25$ but becomes approximately constant for $\cos\theta < 0.1$.  The rate of neutrino-induced muons is expected to be approximately independent of $\cos\theta$~\cite{part}.  We select events with $\cos\theta<0.05$ to minimize the background from cosmic ray muons. To estimate this background we fit  an exponential to the distribution of Fig.~\ref{fig:costheta} for $0.1 \leq \cos\theta \leq 0.2$, a similar procedure to that used by Super-Kamiokande~\cite{Fukuda4}.  The exponential fit is shown in the figure; it has the form $y = \exp(a+b\cos\theta)$, where $a=-3.59$ and $b=51.65$.  Fig.~\ref{fig:costheta} shows that the background at $\cos\theta=0.05$ is 0.3 events and is negligible for smaller values of $\cos\theta$. 

In Table~\ref{table:upHorizontal} we show that there are $\horizontal$ horizontal, neutrino-induced muons in the data sample.  The background shown includes both cosmic ray downward-going muons and the contained vertex events discussed previously.

\begin{table}
\caption{\label{table:upHorizontal}Observed upward-going and horizontal muons. The background accounts for both contained vertex neutrino interactions and downward-going cosmic ray muons.}
\begin{tabular}{|l|c|c|c|} 
\hline \hline
 \multicolumn{1}{|c|}{Muon Type} &~Events~          & \multicolumn{2}{c|}{Background}\\ \hline
 \multicolumn{2}{|c|}{}                                          &~~Contained~~&~~Cosmic $\mu$~~\\ \hline
\hline ~Upward-going~~              & $\upgoing$     & 4.2                       & 0.0\\
\hline ~Horizontal                        & $\horizontal$ &0.1                        & 0.3\\
\hline \hline
\end{tabular}
\end{table}

\subsection{\label{sec:charge}Charge Sign and Momentum Determination}

Previous oscillation analyses based on neutrino-induced muons have typically divided the event sample into through-going and stopping muons.  The experiments that presented these analyses could only measure the muon momentum directly for stopping muons.  Since the MINOS far detector is magnetized, we can measure the muon momentum and charge sign for both types of neutrino-induced muons.

To determine whether the charge and momentum of a neutrino-induced muon has been determined accurately, we have developed a criterion based on the measured curvature of the muon track.  We first draw a straight line between the end points on the track and then use the deviations of the remaining track points from the line to form a $\chi^{2}$ statistic, $\chiLine$.  Large values of $\chiLine$ indicate significant bending in the magnetic field which leads to a good determination of the charge sign and momentum.  

\begin{figure}
\centerline{\epsfig{file=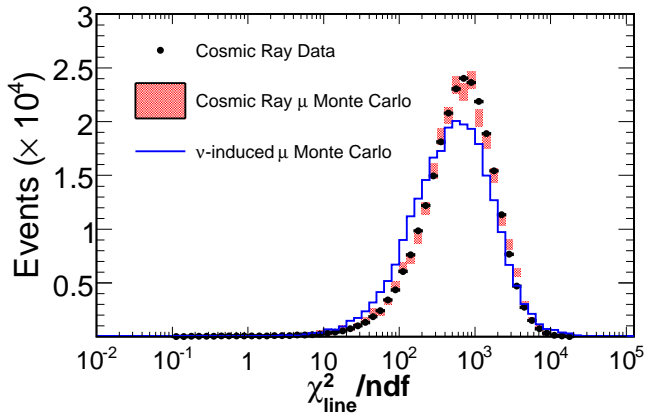, width=3.75in}}
\caption[Data and Monte Carlo distributions of $\chiLine$.]{Comparison of the $\chiLine$ distributions for stopping muons in the cosmic ray data (points), in the cosmic ray Monte Carlo simulation (boxes),  and in the neutrino-induced Monte Carlo simulation (line). The vertical extent of the boxes representing the cosmic ray Monte Carlo simulation indicate the statistical uncertainty for the points.}
\label{fig:chisqrDataMCStop}
\end{figure}

We use our Monte Carlo simulation to set the value of $\chiLine$ that selects neutrino-induced muons with good momentum and charge sign determination.  As a test of whether the $\chiLine$ values are well simulated we compare in Fig.~\ref{fig:chisqrDataMCStop} the distributions of $\chiLine$ for stopping muons in the cosmic ray data, in the cosmic ray muon Monte Carlo simulation, and in the neutrino-induced muon Monte Carlo simulation.  The Monte Carlo distributions have been normalized to have the same number of events as in the data.  Fig.~\ref{fig:chisqrDataMCStop} shows that the cosmic ray muon Monte Carlo simulation provides a high statistics match to the cosmic ray data.  Since the stopping cosmic ray muons have energies $< 20$ GeV, the maximum energy of a normally incident stopping muon, these low energy cosmic events provide a sample of muons with a distribution in energy similar to neutrino-induced muons.  Also superposed on these distributions is that of the Monte Carlo neutrino-induced muons.  This figure suggests that the neutrino-induced Monte Carlo simulation can be used to determine an appropriate value of $\chiLine$ for selecting events with good charge sign and momentum determination.

\begin{figure}
\centerline{\epsfig{file=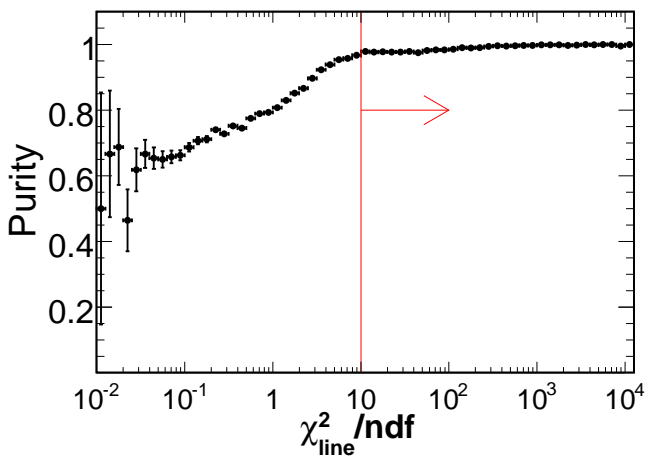, width=3.75in}}
\caption[Charge identification purity as a function of $\chiLine$.]{Purity of the charge sign determination for neutrino-induced muons as a function of $\chiLine$.  Events with $\chiLine>10$ have well-determined charge sign and momentum.}
\label{fig:chisqrPurity}
\end{figure}

In Fig.~\ref{fig:chisqrPurity} we show the fraction of Monte Carlo neutrino-induced muons with correct charge identification, or purity, as a function of $\chiLine$.  The purity is approximately 97\% at $\chiLine = 10$ and rises to over 99\% with increasing values of $\chiLine$.  We identify muons with $\chiLine>10$ as having good charge-sign and momentum determination;  muons with  $\chiLine<10$ are assumed to have poor charge-sign and momentum determination.  

We further divided the Monte Carlo muons with $\chiLine > 10$ into two samples based on their momentum, a separation that roughly distinguishes muons whose parent neutrinos have a relatively large probability of oscillation from those with a lower probability.  The `$\lowMom$' (L) sample includes muons with fit momentum, $p_{fit}$, in the range $1 \leq p_{fit} < 10$ GeV/$c$. The `$\hiMom$' (H) sample includes muons in the range $10 \leq p_{fit} < 100$ GeV/$c$.  No muons with $\chiLine > 10$ have $p_{fit} > 100$ GeV/$c$.  The muons with $\chiLine < 10$ are in the '$\lowRes$' (U) sample.

\begin{figure}
\centerline{\epsfig{file=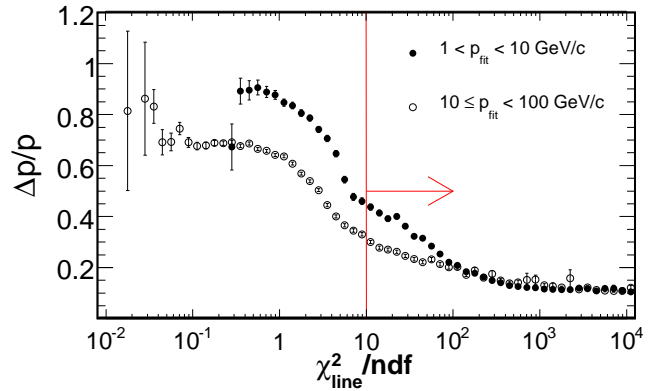, width=3.75in}}
\caption[Momentum resolution as a function of $\chiLine$.]{Momentum resolution for Monte Carlo neutrino-induced muons as a function of $\chiLine$. The line at $\chiLine = 10$ shows the cut used in the analyses.}
\label{fig:chisqrResolution}
\end{figure}

The quality of the momentum determination can be deduced from the Monte Carlo simulation using the r.m.s.~momentum resolution, $(\Delta p/p) = \sqrt{\langle(p_{truth} - p_{fit})^{2}/p_{truth}^{2}\rangle}$, where $p_{truth}$ is the known momentum of the muons.  In Fig.~\ref{fig:chisqrResolution} we show $\Delta p/p$, as a function of $\chiLine$ for the $\lowMom$ and $\hiMom$ samples.  The $\lowMom$ Monte Carlo muons have a momentum resolution of less than 50\% at $\chiLine = 10$ and the resolution decreases to 10\% with increasing values of $\chiLine$.  For the $\hiMom$ muons the momentum resolution is less than 30\% at $\chiLine =10$ and it decreases to 10\% as $\chiLine$ increases.  

\begin{figure}
\centerline{\epsfig{file=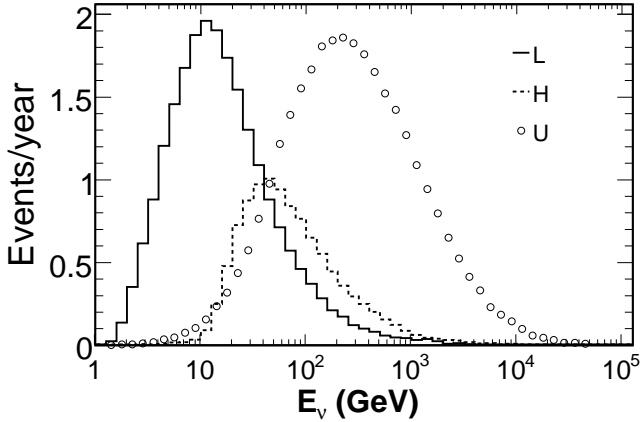, width=3.75in}}
\caption[Distribution of energies for neutrinos producing neutrino-induced muons.]{Distribution of energies for neutrinos producing neutrino-induced muons observed in the MINOS detector as determined by the Monte Carlo simulation. The neutrinos producing $\lowMom$ muons are shown by the solid line, those producing $\hiMom$ muons are shown by the dashed line and those producing $\lowRes$ muons are shown by open circles.  Neutrinos producing muons detected by MINOS have energies $\gtrsim 2$ GeV.}
\label{fig:lowhiparent}
\end{figure}

Fig.~\ref{fig:lowhiparent} shows the distributions of parent neutrino energies for the $\lowMom$, $\hiMom$, and $\lowRes$ muons as determined by the Monte Carlo simulation.  The neutrinos producing $\lowMom$ muons have energies that peak near 10 GeV.  The $\hiMom$ muons are produced by neutrinos with energies that peak near 50 GeV.  The $\lowRes$ muons are produced by even higher energy neutrinos, with a peak energy near 250 GeV.  Comparing the peak energies of the various neutrino distributions we see from Eq.~(\ref{eq:standardprob}) that the $\lowMom$ muons are those that are expected to show the strongest oscillation signal, a result also seen in Fig.~\ref{fig:nuancemom}

\begin{figure}
\centerline{\epsfig{file=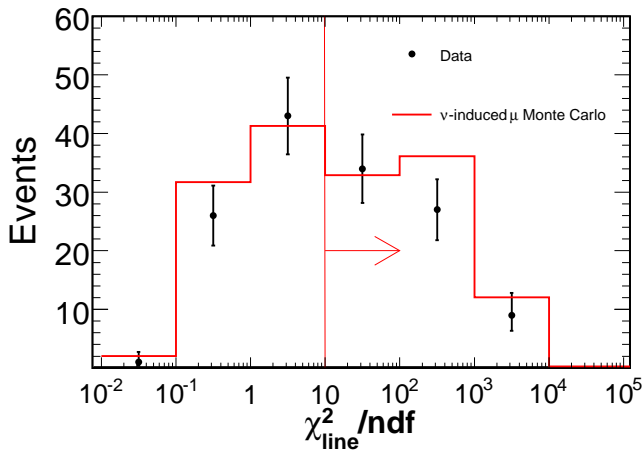, width=3.75in}}
\caption[Data and Monte Carlo distributions of $\chiLine$.]{Comparison of the $\chiLine$ distributions for neutrino-induced muon data and unoscillated Monte Carlo simulation.}
\label{fig:chiSqrDataMC}
\end{figure}

In Fig.~\ref{fig:chiSqrDataMC} we show the distributions of $\chiLine$ for the selected neutrino-induced muons from the data and unoscillated Monte Carlo simulation normalized by live time.  The muons with $\chiLine<10$ are from high energy neutrinos and are not expected to show a strong oscillation signal.  Those muons with $\chiLine>10$ are from lower energy neutrinos,  a sample expected to show an oscillation signal.

\begin{figure}
\centerline{\epsfig{file=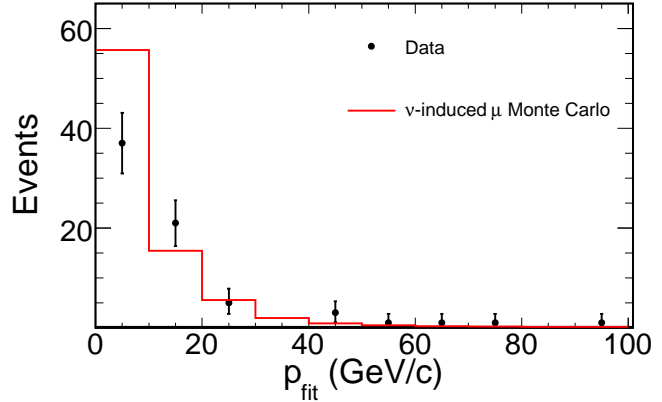, width=3.75in}}
\caption[Distribution of fit momenta.]{Distribution of fit momenta for events in the 
combined $\lowMom$ and $\hiMom$ data samples. The Monte Carlo expectation for no oscillations is shown by the solid line.  The $\lowRes$ muons are not included in this figure.}
\label{fig:abspfit}
\end{figure}

The results of applying the $\chiLine$ cut to the sample of $\candidates$ neutrino-induced muons are given in Table~\ref{table:events}.   In this table, the events with $\chiLine>10$ have been further separated by charge identification and momentum. The calculated background contribution is shown as well as the Monte Carlo expectation in the absence of oscillations.  Figure~\ref{fig:abspfit} shows the fit momentum distributions for data and unoscillated Monte Carlo simulation using the combined $\lowMom$ and $\hiMom$ muon samples.  The $\lowRes$ muons are not included in this figure.  The first bin in Fig.~\ref{fig:abspfit} shows a deficit of events in the data relative to Monte Carlo expectation without oscillations (c.f., Fig.~\ref{fig:nuancemom}).
 
\begin{table}
\caption{\label{table:events}Momentum and charge sign of selected neutrino-induced muons. The calculated background is shown as well as the Monte Carlo expectation in the absence of oscillations.}
\begin{tabular}{|c|c|c|c|}
\hline\hline
\multirow{2}{*}{$p_{fit}$ (GeV)}
&\multirow{2}{*}{~~Data~~}&\multirow{2}{*}{~~Bkgd~~}&\multirow{2}{*}{~~MC~~}\\ 
& & & \\
\hline\hline   & \multicolumn{3}{c|}{$\mum$} \\ \cline{2-4}
$1 - 10$ ~~~~(L)&21  &     2.2    & 37.5 \\
\hline $10 - 100$ ~(H)  &20     & 0.2      & 17.5       \\ 
\hline \hline  &\multicolumn{3}{c|}{$\mup$}\\ \cline{2-4}
$1 - 10$ ~~~~(L)& 16  &  1.3   &  19.3\\
\hline $10 - 100$ ~(H)  & 13  & 0.2        & 8.6      \\ 
\hline \hline
&\multicolumn{3}{c|}{~U} \\\cline{2-4}
unknown~(U)  & 70 & 0.7 &  76.5          \\
\hline \hline
\end{tabular}
\end{table}

\section{\label{sec:results}Oscillation Analysis}

After first discussing the systematic uncertainties associated with the analysis procedure, we test the neutrino-induced muons in Table~\ref{table:events} for evidence of neutrino oscillations.  

\subsection{\label{sec:systematics}Systematic Uncertainties}

There are several sources of systematic uncertainty in this analysis. These are due to both the event reconstruction and the physics modeling.  These uncertainties are summarized in Table~\ref{table:systematics}.

The sources of reconstruction systematic uncertainties are those associated with the data selection cuts (1) $\chi^{2}_{fit}/ndf$, (2) $\chi^{2}_{1/\beta}/ndf$ and (3) $\chiLine$, where the numbers refer to Table~\ref {table:systematics}.  The systematic uncertainties for these cuts were all computed in a similar manner.  For example, we computed the systematic uncertainty on $\chi^{2}_{1/\beta}/ndf$ by establishing that the cut value of $\chi^{2}_{1/\beta}/ndf=3$ selects 98.9\% of events seen in Fig.~\ref{fig:invbetachi2}.  We divided the total Monte Carlo data set into 12 subsamples of 20,000 events each. For each subsample we found the value of $\chi^{2}_{1/\beta}/ndf$ that accepts 98.9\% of the events. The variance in these values of $\chi^{2}_{1/\beta}/ndf$ divided by the nominal value of the $\chi^{2}_{1/\beta}/ndf$ cut is the $1\sigma$ uncertainty given in Table~\ref{table:systematics}.  The $1\sigma$ values for the uncertainties in $\chi^{2}_{fit}/ndf$ and $\chiLine$ were computed in a similar manner.

The first source of systematic uncertainty in the physics modeling is the (4) overall normalization of the calculated neutrino flux.  Uncertainties in the primary cosmic ray fluxes and the hadronic production are the main contributors to the overall uncertainty in the normalization.  Combining these sources gives an uncertainty in the flux normalization of 15\%~\cite{Barr:2006it}.  There are three sources of systematic uncertainty in the ratio of the number of muons induced by low energy neutrinos to those induced by high energy neutrinos.  One contribution comes from (5) the uncertainty in the spectral index of the atmospheric neutrino energy spectrum.  The neutrino flux is proportional to $E_{\nu}^{-(\gamma+1)}$, with the value of the spectral index,  $\gamma=1.7\pm 0.05$~\cite{gaisser}, a 3\% uncertainty in the spectral index.  In addition, the uncertainties in the neutrino and anti-neutrino cross sections contribute another 7\%~\cite{Sanjib2} to the rate of muons coming from neutrinos with energies $< 30$ GeV (6) and 2\%~\cite{PDBook} for muons coming from neutrinos with energies $> 30$ GeV (7).  
  
\begin{table}
\caption{\label{table:systematics}Sources of systematic uncertainty in low to high momentum event ratio ${\cal R}$.}
\begin{tabular}{|l|c|c|} 
\hline\hline
\multicolumn{1}{|c|}{Source}   & ~~~$\sigma$~~~  &  ~$\Delta \mathcal{R}_{k}$~  \\
\hline {\bf Reconstruction systematics: }  &          &  \\
\hline ~~~(1)~$\chi^{2}_{fit}/ndf<1.5$         &  0.01  &  $<5\times10^{-4}$\\
\hline ~~~(2)~$\chi^{2}_{1/\beta}/ndf<3.0$  & 0.01 & $<1\times10^{-4}$ \\
\hline ~~~(3)~$\chiLine<10$~                     & 0.27   &   0.02\\
\hline\hline {\bf Model systematics: }        &          &  \\
\hline   ~~~(4)~Normalization                & 0.15  & $<1\times10^{-4}$ \\
\hline   ~~~(5)~Spectral Index                 & 0.03  & 0.08 \\
\hline   ~~~(6)~Cross Section ($E<30$ GeV) & 0.07  &0.03 \\
\hline   ~~~(7)~Cross Section ($E>30$ GeV) & 0.02    &0.01 \\
\hline \hline $\Delta {\cal R} = \sqrt{\sum ( \Delta \mathcal{R}_k)^2}$ & &0.09 \\
\hline\hline
\end{tabular}
\end{table}

\subsection{\label{sec:lowhiratio}Low to High and Unknown Momentum Event Ratio}

One way to look for evidence of neutrino oscillations in the neutrino-induced muons is to take the ratio of the number of $\lowMom$ muons, which are  more likely to show an oscillation signal, to the sum of the number of $\hiMom$ and $\lowRes$ muons, which are less likely, and compare this ratio with its Monte Carlo expectation including backgrounds.  In the data, this ratio of low to the sum of high and unknown momentum muons is given by
\begin{equation}
R_{{L}/{H+U}}^{data} = \sum_{L} (N_{\mu^-}+N_{\mu^+}){\Large /}
\sum_{H+U} (N_{\mu^-}+N_{\mu^+}),
\end{equation}
where $N_{\mu^{-}}$ is the number of $\mu^{-}$ observed in a bin and $N_{\mu^{+}}$ is the number of $\mu^{+}$ observed in a bin.  The sum over $L$ includes events in the range $1 < p_{fit} < 10$ GeV/$c$, and the sum over $(H+U)$ includes the remaining high momentum and unknown momentum events.  In the Monte Carlo simulation, a similar ratio, $R_{{L}/{H+U}}^{MC}$, is defined.  In the absence of oscillations, the ratio of these two quantities, $\mathcal{R}$, will be consistent with unity; if an oscillation signal is present, $\mathcal{R}$ will be less than unity.  

We computed the systematic uncertainty in this ratio with our Monte Carlo simulation by varying the value of each of the parameters in Table~\ref{table:systematics} by $\pm 1\sigma$ from the values used in the analysis.  The change in the Monte Carlo ratio resulting from the variation in each parameter, $\Delta {\cal R}_k$, was assumed to be the uncertainty in the ratio.   The total uncertainty in $\Delta {\cal R}$  was found by adding the uncertainties from the individual parameters in quadrature.

As the reconstruction uncertainties are based on how different the cut efficiencies are for data and Monte Carlo simulation, we determined how the value of those selection criteria affected the ratio by varying the value of the cuts.  Varying $\chi^{2}_{fit}/ndf$ between 1.485 and 1.515 gives $\Delta \mathcal{R}_1 <5\times10^{-4}$.  The value $\Delta \mathcal{R}_2 <1\times10^{-4}$ is as expected from Fig.~\ref{fig:invbetachi2}. Changing the value of $\chiLine$ between 7.3 and 12.7 shows that $\Delta \mathcal{R}_3 = \pm 0.02$.  The flux normalization simply scales the number of $\lowMom$, $\hiMom$, and $\lowRes$ muons by a constant so this uncertainty cancels in the ratio.  Varying the spectral index by $\pm 3\%$ alters the relative numbers of $\lowMom$ to $\hiMom$ and $\lowRes$ muons which leads to  $\Delta \mathcal{R}_5 = \pm0.08$.  Varying the cross section for neutrinos with $E < 30$ GeV by $\pm 7$\% gives $\Delta \mathcal{R}_6 = \pm 0.03$.  A variation of $\pm 2$\% in the cross section for neutrino interactions with $E > 30$ GeV gives $\Delta \mathcal{R}_7 = \pm 0.01$.  Adding these uncertainties in quadrature gives the total systematic uncertainty, $\Delta\mathcal{R} = 0.09$.  These results are given in Table~\ref{table:systematics}.

From Table~\ref{table:events} and Table~\ref{table:systematics} we find
\begin{equation}
{\cal R} = \frac{R_{{L}/{H+U}}^{data}}{R_{{L}/{H+U}}^{MC}}=\lowHi^{+\lowHiUp}_{-\lowHiDn}(\text{stat})\pm0.09(\text{syst}). 
\label{eq:lowhiresult}
\end{equation}
The upper and lower limits on the data event rate ratio are estimated
accounting for the Poisson fluctuations in the numerator and
demoninator~\cite{Gehrels:1986mj}. The ranges quoted are calculated to
give coverage at 68\% C.~L.  Adding the upper statistical and systematic uncertainties in quadrature, the upper uncertainty is $+0.17$ which results in a value for ${\cal R}$  that differs from the no oscillation expectation of unity by $2.0\sigma$.  This result is consistent with neutrino oscillations.  

\subsection{\label{sec:oscFit}Oscillation Fit}

In the following section we test the significance of the neutrino disappearance suggested by the value of $\mathcal{R}$ in Eq.~(\ref{eq:lowhiresult}) by fitting for the oscillation parameters $\sinSqr$ and $\deltaMSqr$.

\subsubsection{\label{sec:meth}Method}
We used the data and Monte Carlo samples from Table~\ref{table:events} for the oscillation fit.  We first divided the momentum-separated muons (L,H,U) into three $\cos\theta$ bins: $-1.0<\cos\theta<-0.6$, $-0.6<\cos\theta<-0.2$ and $-0.2<\cos\theta<0.05$. To find the oscillation parameters $(\sinSqr,\deltaMSqr)$, we compared  the data and Monte Carlo simulation with the $\chi^{2}$ statistic,
\begin{equation}
\chi^{2} =  2\sum_{i=1}^{9} \biggl(N^{\text{MC}}_{i} - N^{\text{D}}_{i}+ N^{\text{D}}_{i}\ln\frac{N^{\text{D}}_{i}}{N^{\text{MC}}_{i}}\biggr)+\sum_{k=1}^{7}\frac{\epsilon_{k}^{2}}{\sigma_k^{2}}.
\label{eq:chisqrforfit}
\end{equation}
The first sum is the $\chi^{2}$ for Poisson distributed data~\cite{PDBook1}. The number of data events seen in bin $i$ is $N^{\text{D}}_{i}$ and the number of events predicted in bin {\it i} by the Monte Carlo simulation is $N^{\text{MC}}_{i}$.  The parameters $\epsilon_{k}$ in the second sum account for the systematic uncertainties described in \S\ref{sec:systematics}.  The $\epsilon_{k}$ are included in the fit as penalty terms and allow the fit to adjust the predicted number of muons in each bin $i$ based on the systematic uncertainties.  The $\epsilon_{k}$ are assumed to be Gaussian distributed with a mean of 0 so that excursions of the fit values from the expected values contribute an amount $\epsilon_{k}^{2}/\sigma_k^{2}$ to the $\chi^{2}$ statistic, where the $\sigma_{k}$ are given in Table~\ref{table:systematics}.

\begin{figure}
\centerline{\epsfig{file=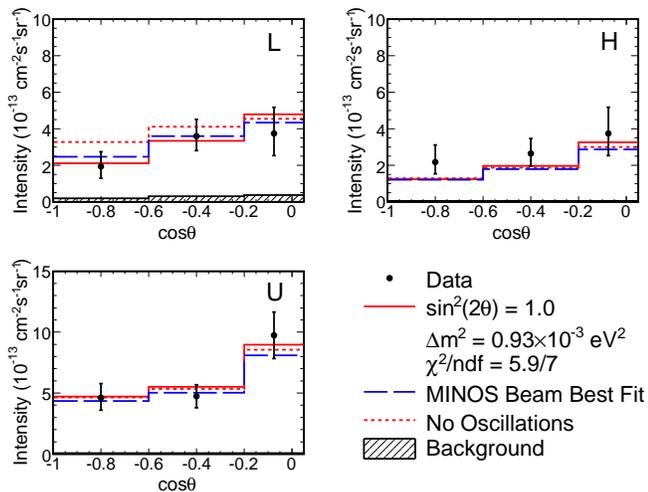, width=3.5in}}
\caption[Result of oscillation fit.]{The intensity of neutrino-induced muons as a function of $\cos\theta$.  The data are shown by the points, the best fit is shown by the solid line, and the null oscillation hypothesis is shown by the dotted line.  The prediction using the MINOS result with the NuMI neutrino beam is shown by the dashed line. The top left panel shows the events with $1 < p_{fit} < 10$ GeV/$c$ (L), the top right shows the events with $10 \leq p_{fit} < 100$ GeV/$c$ (H) and the bottom left shows events with unknown momentum (U).}
\label{fig:costhetaflux}
\end{figure}

The $N^{\text{MC}}_{i}$ depend on the effects of neutrino oscillations, the systematic uncertainties in the reconstruction and input physics model, and the background present for bin $i$.  We define the expected number of muons in bin $i$ as $N^{\text{MC}}_{i} = N^{\text{MC}}_{i}(\sinSqr,\deltaMSqr,\vec\epsilon)+N^{\text{BG}}_{i}$. The background events due to neutrino interactions inside the detector, $N^{\text{BG}}_{i}$, have been oscillated with the best fit Super-Kamiokande parameters and $\vec\epsilon = (\epsilon_{1},\ldots,\epsilon_{7})$ correspond to the uncertainties listed in Table~\ref{table:systematics}.  We calculated the values of $\chi^{2}$ at each point on a grid in $(\sinSqr,\deltaMSqr)$ space.  The fit varied the $\epsilon_{k}$ at each point to find the minimum $\chi^{2}$~\cite{NumRec} in Eq.~(\ref{eq:chisqrforfit})~\cite{Ashie:2005ik}.  

\subsubsection{\label{resOscFit}Results of Oscillation Fit}

The best fit in the physical region is found at $\sinSqr = 1$ and $\deltaMSqr = 0.93\times10^{-3}$ eV$^{2}$ with $\chi^{2}/ndf = 5.9/7$.  The values of the $\epsilon_{k}$ were all found to be less than $1\sigma_{k}$.  The results of the fit are shown in Fig.~\ref{fig:costhetaflux}, where the best fit to the data in the physical region (solid line), the Monte Carlo prediction for the null oscillation hypothesis (dotted line), and the prediction derived from the MINOS result with the NuMI neutrino beam (dashed line)~\cite{MINOSnu} are superposed onto the observed intensity of neutrino-induced muons.  For the results from the null oscillation hypothesis and the MINOS beam analysis we varied the $\epsilon_{k}$ to minimize Eq.~(\ref{eq:chisqrforfit}) at the fixed values of $\sinSqr$ and $\deltaMSqr$.

\begin{figure}
\centerline{\epsfig{file=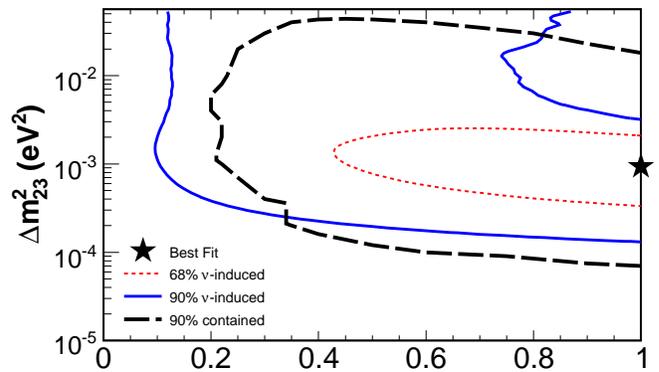, width=3.75in}}
\caption[Confidence intervals for the oscillation fit.]{The the 68\% (dotted line) and 90\% (solid line) confidence intervals for the oscillation parameter fit.  The best fit point is indicated by the star.  Also shown is the 90\% confidence interval for the MINOS contained vertex analysis (dashed line) based on the first 418 days of data taking with the far detector.}
\label{fig:contours}
\end{figure}

Fig.~\ref{fig:contours} shows the 68\% and 90\% confidence intervals for the oscillation parameter fit.  The points within the 68\% contour have a difference in $\chi^{2}$ with the best fit point, $\Delta\chi^{2}$, of less than 2.3.  The points within the 90\% contour have $\Delta\chi^{2}<4.61$.  The null oscillation hypothesis has $\Delta\chi^{2}=5.5$ and is excluded at the $\nullExclude$\% confidence level.  Also shown in Fig.~\ref{fig:contours} is the 90\% confidence interval for the MINOS contained vertex analysis~\cite{Adamson:2005qc} based on the first 418 days of data taking with the far detector.  The results of these two analyses are consistent.

\section{\label{sec:CPT}Tests of CPT Conservation}

To probe CPT conservation with MINOS data, we use two tests.  After discussing the systematic uncertainties, we first compare the strength of the oscillation signals for the neutrino-induced $\mu^-$ and $\mu^+$ samples individually.   Second, we divide the charge-separated muons into low and high momentum samples and further test for differing rates of neutrino disappearance.  We show that this second test is the one that minimizes systematic uncertainties.

\subsection{\label{sec:systematicsCPT}Systematic Uncertainties}

Here we discuss the additional sources of systematic uncertainties associated with the CPT analysis.  One additional source of systematic uncertainty is due to misidentified charge sign or the charge purity of the data sample.   We determined the charge purity systematic by comparing the underground cosmic ray muon charge ratio for events passing the $\chiLine$ cut with the charge ratio found by MINOS, $N_{\mu^{+}}/N_{\mu^{-}} = 1.371$~\cite{m&r}.   For this comparison we divided the cosmic ray muons into two data samples, one with muons having momenta $< 30$~GeV/$c$ and the other with momenta in the range 30~--~100~GeV/$c$.  For cosmic ray muons in the lower momentum sample, we calculated the difference in the charge ratio to be 0.047,  a difference that can be attributed to an impurity of 5\%.  For cosmic ray muons in the high momentum sample, we calculated the difference in the charge ratio to be 0.12, which can be attributed to an impurity of 14\%.  By taking a weighted average of these impurities, where the weights are given by the number of neutrino-induced muons in each momentum range, we calculated the impurity in the neutrino-induced muon data to be 6\%.  Using similar methods we found the weighted average impurity for the neutino-induced Monte Carlo events to be 1\%.  These weighted impurities are given in Table~\ref{table:cptsystematics3} (4). 

There are also additional sources of systematic uncertainty due to the physics model.  These arise from uncertainties in the ratios of the $\numbar$ cross section, $\sigma_{\numbar}$, to the $\num$ cross section, $\sigma_{\num}$, and we estimate the uncertainties to be 9\%~\cite{Sanjib2} for energies $<30$ GeV~(9) and 2\%~\cite{PDBook2} for energies $>30$ GeV~(10).  The uncertainty in the ratio of the number of $\num$ to $\numbar$~(11) is 1\%~\cite{Barr:2006it}.

\subsection{\label{sec:CPTRatio}Charge-Separated Event Ratio}

Our first test of CPT conservation comes from the comparison of the ratio of the total number of $\mu^-$ events to $\mu^+$ events in the data and the Monte Carlo simulation including backgrounds.  In the data, this ratio of charge separated muons is given by
\begin{equation}
R^{data}_{\mu^-/\mu^+}  = \sum (N_{\mu^-}){\Large /} \sum (N_{\mu^+}).
\label{eq:rdata1}
\end{equation}
The Monte Carlo ratio is defined in a similar way.  If the $\num$ oscillate with the same parameters as the $\numbar$ then the ratio of the data and Monte Carlo ratios, $\hat{\mathcal{R}}_{CPT}$, will be consistent with unity.  The systematic uncertainties in $\hat{\mathcal{R}}_{CPT}$ are shown in Table~\ref{table:cptsystematics3}.  This table shows how changes in the default parameters corresponding to the sources of systematic uncertainty changes the value of $\hat{\mathcal{R}}_{CPT}$.  The main contributions to the uncertainty in  $\hat{\mathcal{R}}_{CPT}$ are the purity, cross section ratio for neutrinos with energies $<30$ GeV, and the ratio of $\num/\numbar$.  Since the ratio in Eq.~(\ref{eq:rdata1}) is greater than one, more negative than positive muons will be misidentified and the measured charge ratio decreases towards unity.  Consequently, charge misidentification leads to a one-sided (positive) error on the measured charge ratio. The uncertainty due to the impurity is $(\Delta \hat{\mathcal{R}}_{CPT})_{4} = 0.07$. The cross section ratio for neutrinos with energies $<30$ GeV changes the ratio by $(\Delta \hat{\mathcal{R}}_{CPT})_{9} = \pm 0.04$.   The uncertainty in the ratio of the number of $\num$ to $\numbar$ corresponds to $(\Delta \hat{\mathcal{R}}_{CPT})_{11} = \pm 0.01$. We find from Table~\ref{table:events} and Table~\ref{table:cptsystematics3}
\begin{equation}
\hat{\mathcal{R}}_{CPT} = \frac{R^{data}_{{\mu^-}/{\mu^+}}}{R^{MC}_{{\mu^-}/{\mu^+}}} = \mtop^{+\mtopUp}_{-\mtopDn}(\text{stat})\mtopSys(\text{syst}),
\label{eq:totalRatio}
\end{equation}
where the statistical uncertainties were calculated using the method for Eq.~(\ref{eq:lowhiresult}). This value of $\hat{\mathcal{R}}_{CPT}$ is consistent with  CPT conservation.

\begin{table}
\caption{\label{table:cptsystematics3}Sources of systematic uncertainty considered and their effects on the ratio $\hat{\mathcal{R}}_{CPT}$.}
\begin{tabular}{|l|c|c|} 
\hline\hline
& & \\ 
\multicolumn{1}{|c|}{Source}& ~~$\sigma$~~ & $(\Delta\mathcal{\hat{R}}_{CPT})_k$\\
\hline\hline {\bf Reconstruction systematics}:   &    \multicolumn{2}{c|}{  }  \\
\hline ~~~(1)~$\chi^{2}_{fit}/ndf<1.5$           & 0.01  & $<5\times10^{-4}$\\
\hline ~~~(2)~$\chi^{2}_{1/\beta}/ndf<3.0$    & 0.01 & $<1\times10^{-4}$\\
\hline ~~~(3)~$\chiLine<10$ ~                      & 0.27 & $<5\times10^{-3}$ \\
\hline ~~~(4)~Charge Purity (Data,MC)~       & (0.06,0.01) & +0.07  \\ 
\hline\hline {\bf Model systematics}:    &    \multicolumn{2}{c|}{  }  \\
\hline ~~~(5)~Normalization                           & 0.15 & $<1\times10^{-4}$\\
\hline ~~~(6)~Spectral Index                            & 0.03 & $<1\times10^{-4}$ \\ 
\hline ~~~(7)~Cross Section ($E<30$ GeV)       & 0.07 & $<5\times10^{-3}$\\  
\hline ~~~(8)~Cross Section ($E>30$ GeV)       & 0.02 & $<5\times10^{-3}$\\
\hline ~~~(9)~$\sigma_{\numbar} / \sigma_{\num}$ ($E<30$ GeV) & 0.09 & 0.04\\ 
\hline ~~(10)~$\sigma_{\numbar} / \sigma_{\num}$ ($E>30$ GeV) & 0.02 & $<5\times10^{-3}$\\ 
\hline ~~(11)~$\num/\numbar$ & 0.01 & 0.01\\
\hline\hline$ \Delta \mathcal{\hat{R}}_{CPT} = \sqrt{ \sum (\Delta\mathcal{\hat{R}}_{CPT})_k^2}$   &   & -0.04, +0.08 \\
\hline\hline
\end{tabular}
\end{table}

\subsection{\label{sec:CPTeventRatio}Charge-Separated, Low-to-High Momentun Event Ratio}

Our second test of CPT conservation is based on the charge separated ratios of $\lowMom$ to $\hiMom$ events compared to the Monte Carlo expectation including backgrounds.  First define ratios of data to Monte Carlo expectation for the $\num$ and $\numbar$ individually, 
\begin{eqnarray}
{\cal R}_{-} & = & \biggl(\frac{R^{data}_{{L}/{H}}}{R^{MC}_{{L}/{H}}}\biggr)_{\mu^-} 
\nonumber \\
                         & = & \lowHiM^{+\lowHiMUp}_{-\lowHiMDn}(\text{stat})\pm0.07(\text{syst}),  
\label{eq:cptresult1}
\end{eqnarray}
\noindent and 
\begin{eqnarray}
{\cal R}_{+}& = & \biggl(\frac{R^{data}_{{L}/{H}}}{R^{MC}_{{L}/{H}}}\biggr)_{\mu^+}
\nonumber \\
                           &=&\lowHiP^{+\lowHiPUp}_{-\lowHiPDn}(\text{stat})\pm0.07(\text{syst}). 
\label{eq:cptresult2}
\end{eqnarray}
The systematic uncertainties in these ratios are listed in Table~\ref{table:cptsystematics2}, and the statistical uncertainties were again calculated using the same method as for Eq.~(\ref{eq:lowhiresult}).  We define the ratio of ratios $\tilde{\mathcal{R}}_{CPT}$,
\begin{equation}
\tilde{\mathcal{R}}_{CPT}= {\cal R}_{-}/ {\cal R}_{+} =\cpt^{+\cptUp}_{-\cptDn}(\text{stat})\pm\cptSys(\text{syst}). 
\label{eq:cptresult}
\end{equation}
We found the upper and lower limits for the statistical uncertainty in Eq.~(\ref{eq:cptresult}) by fitting the observed L and H event rates for $\mum$ and $\mup$ to a model which used $\tilde{\mathcal{R}}_{CPT}$ and three of the four event rates as inputs. The data and model were compared using a $\chi^{2}$ function appropriate for Poisson distributed data~\cite{PDBook1}. The 68\% C.~L.~interval quoted includes all values of $\tilde{\mathcal{R}}_{CPT}$ which produce a $\chi^{2}$ within 1 of the best fit value.  Again within the statistical and systematic uncertainties, the value of $\tilde{\mathcal{R}}_{CPT}$ is consistent with unity and CPT conservation.  

The value of this test is clearly demonstrated in Table~\ref{table:cptsystematics2} where it can be seen that several sources of systematic uncertainty cancel in the calculation of $\tilde{\mathcal{R}}_{CPT}$.  These cancellations occur because the $\mum$ and $\mup$ ratios are affected similarly for these systematic uncertainties and so divide out in the ratio.  The uncertainty in the ratio of the number of $\num$ to the number of $\numbar$ does not contribute to the uncertainty in $\tilde{\mathcal{R}}_{CPT}$.  As seen in Table~\ref{table:cptsystematics2}, the only sources of uncertainty which do not cancel are those in the relative magnitude of the $\num$ and $\numbar$ cross sections at low (9) and high (10) energies.  

This approach should prove valuable in high statistics accelerator-based tests of CPT conservation because it minimizes systematic uncertainties so effectively.

\begin{table*}
\caption{\label{table:cptsystematics2}Sources of systematic uncertainty considered and their effects on the ratios $\mathcal{R_{-}}$, $\mathcal{R_{+}}$, and $\mathcal{\tilde{R}_{CPT}}$.}
\begin{tabular}{|l|c|c|c|c|c|c|} 
\hline\hline
& & \multicolumn{2}{c|}{ } & \multicolumn{2}{c|}{ }&\\ 
\multicolumn{1}{|c|}{Source}& ~~$\sigma$~~ & \multicolumn{2}{c|}{$\Delta\mathcal{R}_{-}$} & \multicolumn{2}{c|}{$\Delta\mathcal{R}_{+}$} & 
$(\Delta {\mathcal {\tilde{R}}_{CPT}})_k$\\
\hline & & $-1\sigma$ & $+1\sigma$& $-1\sigma$&$+1\sigma$ &\\
\hline\hline {\bf Reconstruction systematics}:   &    \multicolumn{6}{c|}{  }  \\
\hline ~~~(1)~$\chi^{2}_{fit}/ndf<1.5$           &  0.01 & $<5\times10^{-4}$ & $<5\times10^{-4}$ & $<5\times10^{-4}$   & $<5\times10^{-4}$ & --\\
\hline ~~~(2)~$\chi^{2}_{1/\beta}/ndf<3.0$    &  0.01 & $<1\times10^{-4}$ & $<1\times10^{-4}$ & $<1\times10^{-4}$ & $<1\times10^{-4}$ & --\\
\hline ~~~(3)~$\chiLine<10$ ~                      & 0.27 & 0.05& 0.05  & 0.05 & 0.05 & -- \\
\hline ~~~(4)~Charge Purity  (Data,MC)        & (0.06,0.01) & $<1\times10^{-4}$ & $<1\times10^{-4}$ &  $<1\times10^{-4}$ & $<1\times10^{-4}$ & -- \\
\hline\hline {\bf Model systematics}:    &    \multicolumn{6}{c|}{  }  \\
\hline ~~~(5)~Normalization                           & 0.15 & $<1\times10^{-4}$  &  $<1\times10^{-4}$ & $<1\times10^{-4}$ & $<1\times10^{-4}$ & --\\
\hline ~~~(6)~Spectral Index                            & 0.03 &  0.04  &  0.03 & 0.04 & 0.04 & -- \\
\hline ~~~(7)~Cross Section ($E<30$ GeV)       & 0.07 & 0.02   & 0.02 & 0.02  & 0.02 & --\\
\hline ~~~(8)~Cross Section ($E>30$ GeV)       & 0.02 & 0.01  & 0.01 & 0.01 & 0.01  & --\\
\hline ~~~(9)~$\sigma_{\numbar} / \sigma_{\num}$ ($E<30$ GeV) & 0.09 & --      & --     & 0.02  & 0.02 & 0.03\\
\hline ~~(10)~$\sigma_{\numbar} / \sigma_{\num}$ ($E>30$ GeV) & 0.02 & --      & --     & 0.01  & 0.01 & 0.01\\
\hline\hline  $\Delta {\mathcal {\tilde R}}_{CPT} = \sqrt{ \sum ( \Delta {\mathcal {\tilde R}}_{CPT})_k^2}$   &    \multicolumn{5}{c|}{  }  & $\cptSys$ \\
\hline\hline
\end{tabular}
\end{table*}

\section{Conclusions}
The MINOS far detector has taken $\liveTime$ days of data in its search for the atmospheric neutrino-induced muons.  We found a total of $\candidates$ neutrino-induced muons in the data set.  We have performed four analyses.  The ratio of the number of $\lowMom$ muons to the sum of $\hiMom$ and $\lowRes$ muons has been compared to the same ratio for Monte Carlo events.  The resulting ratio, ${\cal R}$, suggests neutrino disappearance at the $2.0\sigma$ level.  The best oscillation fit to the data gives $\sinSqr = 1.0$ and $\deltaMSqr = 0.93\times10^{-3}$ eV$^{2}$.  This fit excludes the null oscillation hypothesis at the $\nullExclude$\% confidence level.  We used two methods to look for evidence of different rates of neutrino disappearance as a test of CPT conservation.  Both tests are consistent with CPT conservation.  The cancellation of many systematic uncertainties in the computation of $\tilde{\mathcal{R}}_{CPT}$  suggests that this test would provide a precision test of CPT conservation with a suitably large dataset.

\begin{center}
{\bf Acknowledgments}
\end{center}

We thank the Fermilab staff and the technical staffs of the participating institutions for their vital contributions.  This work was supported by the U.S. Department of Energy, the U.S. National Science Foundation, the U.K. Particle Physics and Astronomy Research Council, the State and University of Minnesota, the Office of Special Accounts for Research Grants of the University of Athens, Greece, FAPESP (Rundacao de Amparo a Pesquisa do Estado de Sao Paulo), CNPq (Conselho Nacional de Desenvolvimento Cientifico e Tecnologico) in Brazil, and the computational resources of the AVIDD cluster at Indiana University.  We gratefully acknowledge the Minnesota Department of Natural Resources for their assistance and for allowing us access to the facilities of the Soudan Underground Mine State Park.  We also thank the crew of the Soudan Underground Physics laboratory for their tireless work in building and operating the MINOS detector.

\bibliography{upmuPRD}

\end{document}